\documentclass{PoS}

\title{Pion physics on the lattice}

\ShortTitle{Pion physics on the lattice}

\author{\speaker{Sinya Aoki}
\\
        Graduate School of Pure and Applied Sciences, University of Tsukuba, Tsukuba, Ibaraki 305-8571, Japan\\
        E-mail: \email{saoki@het.ph.tsukuba.ac.jp}}

\abstract{Recent results on pion physics from lattice QCD are reviewed. 
We discuss quark mass dependences of pion mass and decay constant and compare them with  the predictions from chiral perturbation theory. In particular we focus on the convergence of chiral perturbation theory around strange quark mass region.
We also consider  quark mass as well as momentum dependences of pion form factors in recent full QCD simulations. }

\FullConference{6th International Workshop on Chiral Dynamics, CD09\\
		July 6-10, 2009\\
		Bern, Switzerland}

\begin{document}

\section{Introduction}
Nowadays lattice QCD and chiral perturbation theory rely on each other for their developments.
In lattice QCD simulations, besides statistical errors, there exist several systematic errors such as finite size effect and finite lattice spacing effect.
The effect due to the heavier $u,d$ quark masses is one of such systematic errors:
Typical $u,d$ quark masses employed in current lattice QCD simulations are heavier than their physical values, and one has to make extrapolations of results to the physical quark masses using, for example,
chiral perturbation theory(ChPT). Finite size effects in lattice QCD simulations  may also be corrected by ChPT.  On the other hand, lattice QCD provide useful tools to check the convergence of ChPT and to determine low energy constants(LECs) of ChPT, by  varying quark masses in simulations 

In this review, we compare results related to pion physics obtained from various lattice QCD simulations
with predictions from ChPT. We mainly consider mass and decay constant of pion. In addition we briefly discuss recent results of pion form factors for full QCD simulations and compare both momentum and quark mass dependences of form factors with predictions by ChPT.
In this review we will not collect and compare values of LECs from various simulations and groups, since an excellent review on LECs of ChPT from lattice QCD has already existed\cite{necco}.

\section{Recent full QCD simulations}
Both increases of computational resources and improvements of numerical algorithms
enable us to perform full QCD simulations with vert light quark masses.
The lattice spacing $a$(fm), the lattice size $L$(fm), the minimum pion mass $m_\pi^{\rm min}$(MeV) and $L m_\pi^{\rm min}$ for recent large scale full QCD simulations are listed in table 
\ref{tab:fullQCD}, 
in the upper half of which, simulations with the conventional quark action such as Wilson-type or staggered-type quark action are collected, while those with the chiral symmetric action such as the overlap or domain-wall quark action are given in the lower half.
\begin{table}[tb]
\caption{(Incomplete) list of recent full QCD simulations.}
\label{tab:fullQCD}
\vskip -0.2cm
\begin{center}
\begin{tabular}{|c|c|c|c|c|}
\hline
\hline
Group & $a$(fm) & $L$(fm) & $m_\pi^{\rm min}$(MeV) & $ L m_\pi^{\rm min}$ \\
\hline
\hline
\multicolumn{5}{| l |}{Conventional quark action} \\
\hline
2+1 flavors & & & &\\ 
\hline
PACS-CS\cite{pacs-cs}  & 0.09 & 2.9 & 160 &2.3 \\
MILC\cite{milc}  & $\ge 0.06$ & 3.3 & 240 & 4\\
BMW\cite{bmw}  & $\ge 0.065$ & $ \ge 4.2$ & 190 & 4 \\
J-Lab.\cite{jlab}  & $ 0.012$ & $ 1.5-2.9$ & 385 & 5.7 \\
\hline
2 flavors & & & & \\
\hline
CERN-ToV\cite{cern}&   $\ge 0.05$ & 1.7-1.9 & 300& 2.9 \\
ETMC\cite{etmc} & $\ge0.07$ & 2.1 & 300 &  3.2  \\
CLS\cite{cls} & $ 0.08$ & 2.6 & 230 & 3   \\
QCDSF\cite{qcdsf} & $\ge 0.072$ & $2.3$ & 240 & 2.8  \\  
\hline
\hline
\multicolumn{5}{| l |}{Chirally symmetric quark action}\\
\hline
2+1 flavors & & & &  \\
\hline
RBC-UKQCD\cite{rbc-ukqcd}   & 0.11 & 2.8 & 330 & 4.6\\
JLQCD\cite{jlqcd_nf3}   & 0.11 & 1.8 & 315 & 2.8\\
\hline
2 flavors & & & & \\
\hline
RBC\cite{rbc}  & 0.12 & 2.5 & 490 & 6.1\\
 JLQCD\cite{jlqcd_nf2}  & 0.12 & 1.9 & 290& 2.8\\
\hline
\hline
\end{tabular}
\end{center}
\vskip -0.7cm 
\end{table}

\begin{figure}[bt]
\centering
\includegraphics[width=57mm, angle=270,clip]{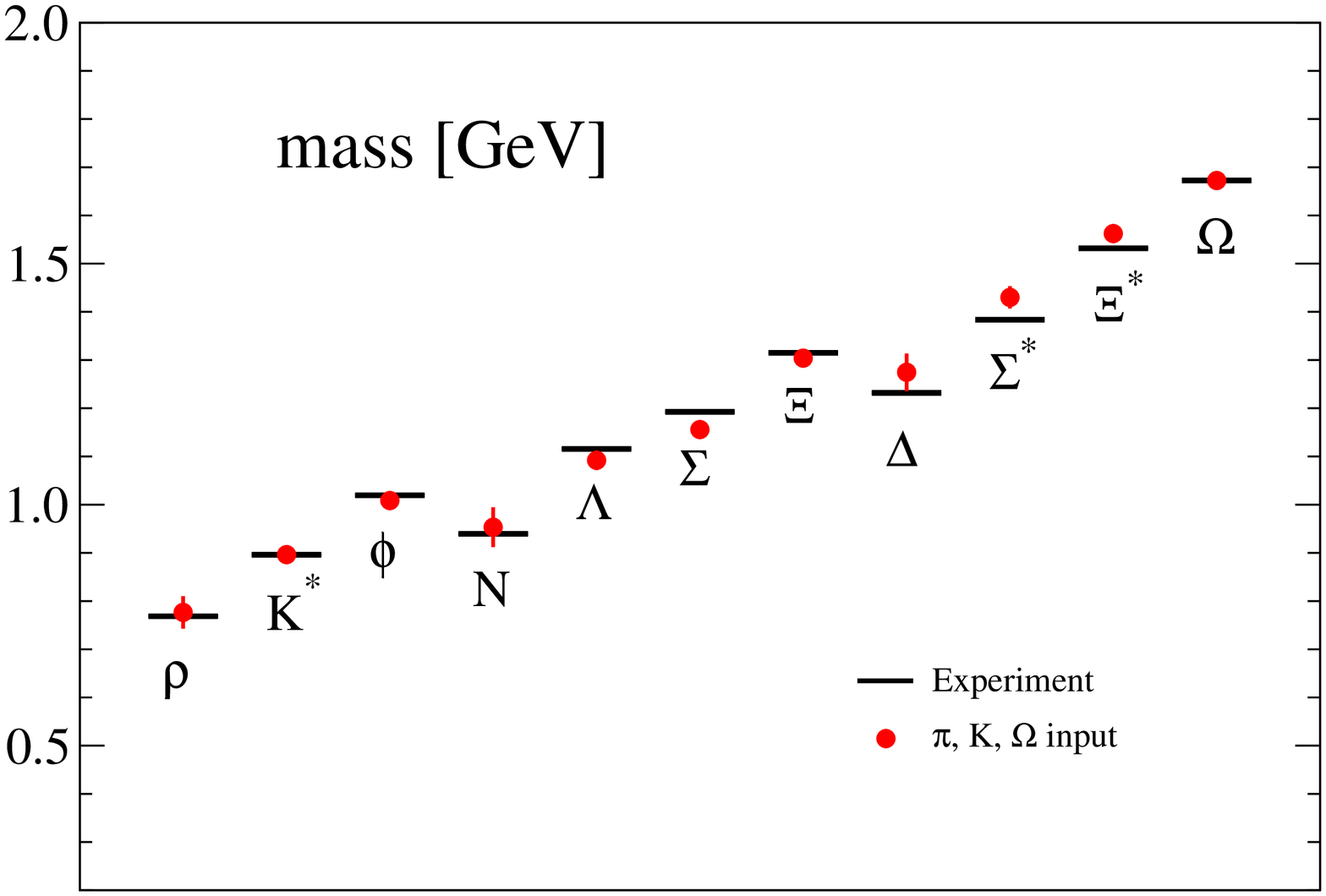}
\includegraphics[width=57mm, angle=270,clip]{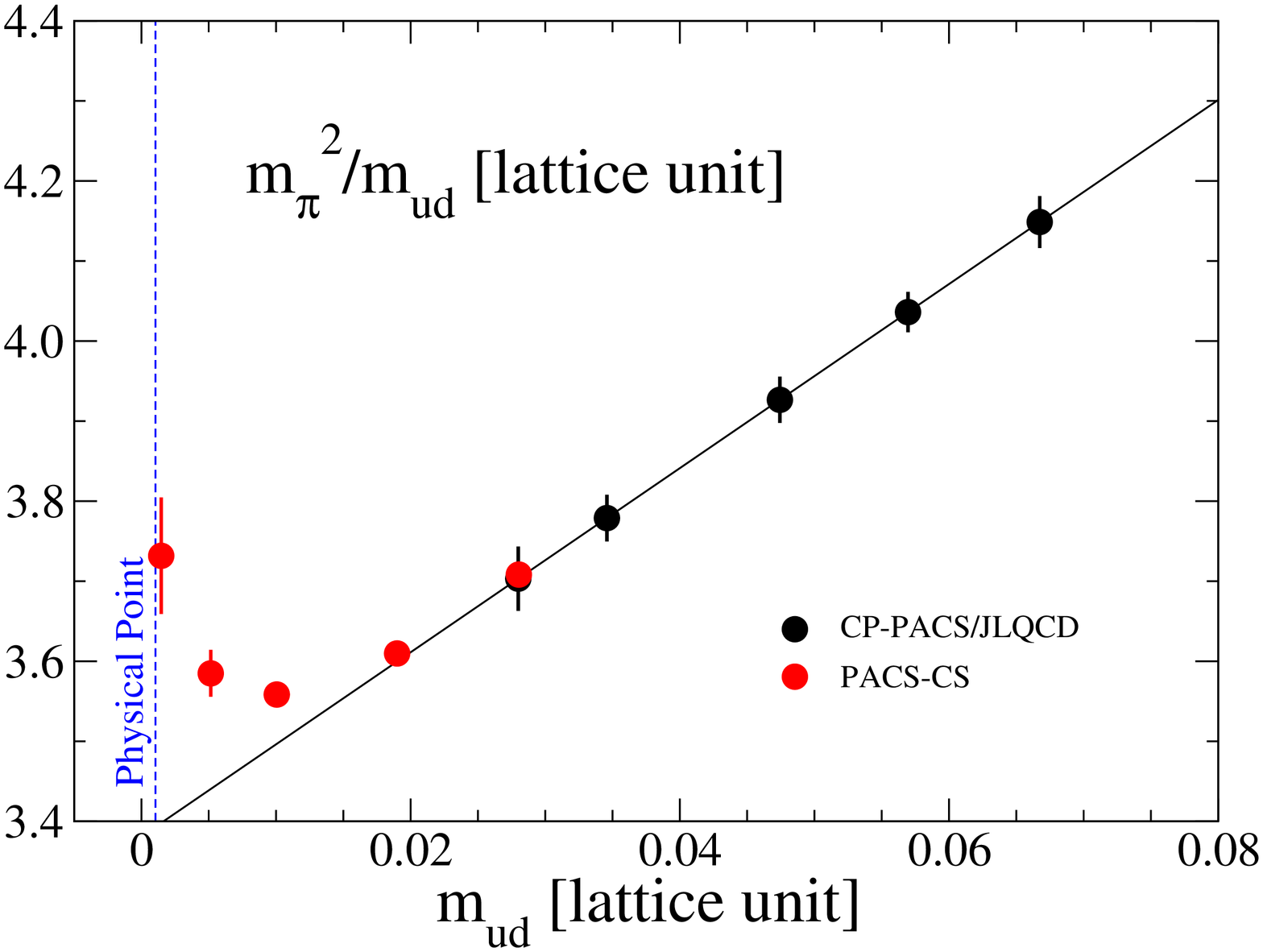}
\caption{Left: Hadron spectrum at $a=0.09$ fm\cite{pacs-cs} together with experimental values. Inputs to fix $m_{ud}$, $m_s$ and $a$ in QCD are $m_\pi$, $m_K$ and $m_\Omega$, respectively.
Right: The ratio $m_\pi^2/m_{ud}$ as a function of $m_{ud}$ at $a=0.09$ fm and $L=2.9$ fm\cite{pacs-cs}. The 5 data points lowest in mass(PACS-CS) correspond to $m_\pi =702,570, 412, 296$ and 156 MeV (from right to left). The vertical dashed line is the physical point. }
\label{fig:pacs-cs}
\vskip -0.4cm
\end{figure}

In Fig.\ref{fig:pacs-cs}(left), one of the recent results for hadron spectra from PACS-CS collaboration\cite{pacs-cs} is presented. This result is obtained at $a=0.09$ fm and $L=2.9$ fm
with 2+1 flavor full QCD using the non-perturbatively $O(a)$ improved Wilson quark action.
Hadron spectra, where masses of $\pi$, $K$ and $\Omega$ are used to fix the light quark mass, the strange quark mass and the lattice spacing,   agree reasonably well with experimental values, even though the continuum extrapolation has not been taken yet.
The minimum pion mass in this simulation reaches 156 MeV, which is almost equal to the physical pion mass, though the finite size effect could be large at $m_\pi L = 2.3$ of this pion mass.
In  Fig.\ref{fig:pacs-cs}(right), the pion mass squared in this simulation divided by the light quark mass ( $m_\pi^2/m_{ud}$) is plotted as a function of the light quark mass. 
As the quark mass decreases, $m_\pi^2/m_{ud}$ positively deviates from the linear behavior
seen in the previous result of PACS/JLQCD collaborations at heavier quark masses.
This deviation is expected form chiral perturbation theory(ChPT) at the next-to-leading order(NLO).
To check a magnitude of a possible finite size effect on the spectra at the lightest quark mass, a new simulation at $L=5.8$ fm and $m_\pi \simeq 140$ MeV ($m_\pi L > 4$), which is indeed considered as a "real QCD simulation",  is on-going\cite{kuramashi}.
    
\section{Pion mass and decay constant and chiral perturbation theory}
In this section we compare quark mass dependences of pion mass and decay constant from recent full QCD simulations with predictions by ChPT.
 
\subsection{2-flavor QCD and SU(2) ChPT}
We first consider lattice QCD simulations with 2 degenerate dynamical quarks.
In this case, SU(2) ChPT at NLO predicts\cite{GL}
\begin{eqnarray}
\frac{m_\pi^2}{m_q} &=& 2B\left\{1+\frac{2B m_q}{16\pi^2 f^2}\left[\ln \left(\frac{2B m_q}{\mu^2}\right)-\ell_3(\mu)\right]\right\}
\label{eq:mpi_su2} \\
f_\pi &=& f\left\{1-\frac{2B m_q}{8\pi^2 f^2}\left[\ln \left(\frac{2B m_q}{\mu^2}\right)-\ell_4(\mu)\right]\right\}
\label{eq:fpi_su2}
\end{eqnarray}
where $m_\pi$ is the pion mass, $f_\pi$ is the pion decay constant with the $f_\pi=132$ MeV
normalization, $B$ and $f$ are low energy constants (LECs) at the leading order (LO), and $\ell_{3,4}(\mu)$ are those at NLO with renormalization scale $\mu$.

\begin{figure}[hbt]
\centering
\includegraphics[width=70mm, angle=0,clip]{Figs/JLQCD_mpfp_vp_nlo_3pts.eps}
\includegraphics[width=55mm, angle=0,clip]{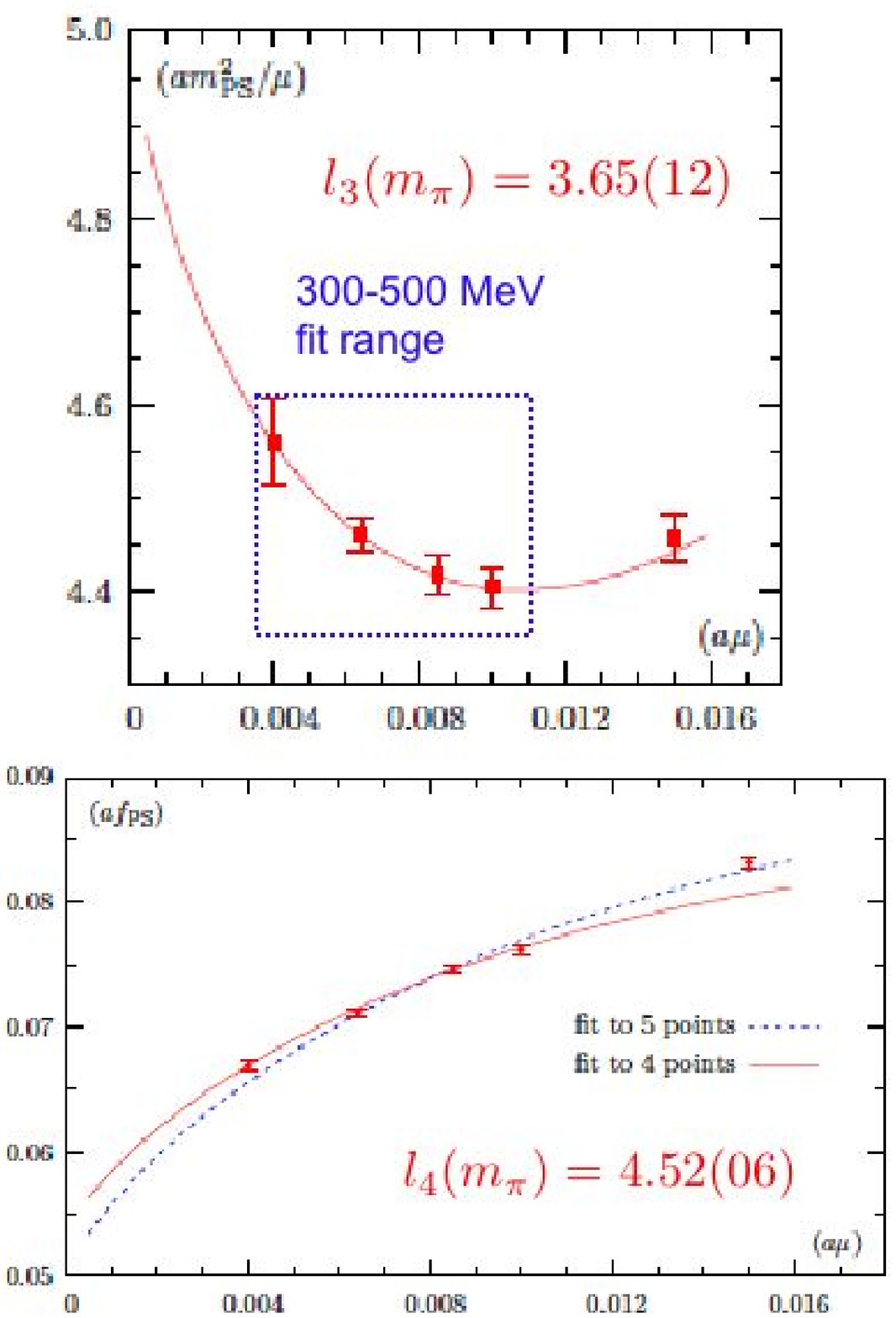}
\caption{Left: $m_\pi^2/m_q$(top) and $f_\pi$ (bottom) from the overlap quark\cite{noaki} as a function of $m_\pi^2$ at $a=0.12$ fm, together with the NLO ChPT fits with 3 different expansion parameters. Right: The same quantities from the twisted mass QCD, taken from Ref.\cite{etmc2}. }
\label{fig:nf2}
\end{figure}

JLQCD and TWQCD collaborations have performed 2-flavor QCD simulations, using the overlap quark,
which has an exact "lattice chiral" symmetry, at $a=0.12$ fm and $L=1.9$ fm with the global topological charge being fixed to $Q=0$\cite{jlqcd_nf2}.  The minimum pion mass in the simulation is $m_\pi = 290$ MeV, which corresponds to $m_\pi L = 2.9$. Before fitting data with ChPT formula, 
$1/V$ finite size effect due to the fixing topological charge\cite{AF} as well as the ordinary exponential type finite size effect\cite{CDH} have been corrected  by ChPT at NLO with phenomenological values for NLO LECs\cite{noaki}. It is found that two finite size corrections tend to cancel each other, so that a total correction is small for both $m_\pi$ and $f_\pi$.
Three different expansion parameters $x=2B m_q/(8\pi^2 f^2)$, $\hat x = m_\pi^2/(8\pi^2 f^2)$ and 
$\xi = m_\pi^2/(8\pi^2 f_\pi^2)$ are employed in the ChPT fit at NLO, 
where $m_\pi$ and $f_\pi$ are measured value of pion mass and decay constant at each quark mass $m_q$. Fig.\ref{fig:nf2}(Left), where $m_\pi^2/m_q$ and $f_\pi$ are plotted as function of $m_\pi^2$,
indicates that  NLO ChPT fits work reasonably well at the lightest 3 pion masses at $m_\pi \le 450$ MeV, for all 3 choices of the expansion parameter. This fact establishes the validity of the NLO ChPT fits
at small enough pion mass ($m_\pi\le 450$ MeV). Furthermore, it is noticeable that the $\xi$-fit
describes data beyond the fitted region. 
The NNLO ChPT fit\cite{CGL} with the $\xi$ is found to be reasonable for all data points at $m_\pi \le 750$ MeV, if a combination of NLO LECs appeared at NNLO is fixed to the phenomenological value. 
However fits show that the NNLO correction become
 30\% for $m_\pi$ and 70\% for $f_\pi$ of the NLO correction at $m_\pi = 500$ MeV, and
variations of some LECs from NLO to NNLO are significant. 

The twisted mass formulation of lattice QCD is defined by a combination of 
the Wilson quark action and the twisted mass term given by
$m_q \bar q(x) e^{i\theta \gamma_5 \tau_3} q(x) $ where $q(x) =(u(x), d(x) )$.
A remarkable property of this formulation is that
$O(a)$ lattice artifacts appeared in the Wilson formulation are absent for physical observables at the maximal twist angle, $\theta_R =\pi/2$, where $\theta_R$ is a renormalized twist angle. 
European Twisted Mass Collaborations (ETMC) have employed the twisted mass lattice QCD at the maximal twist  for 2-flavor full QCD simulations, by numerically tuning $\theta_R\simeq \pi/2$, at
$a=0.087$,  0.067 fm and $L\simeq 2.1$ fm\cite{etmc,etmc2,etmc3}. The minimum pion mass is 310 MeV, so that $m_\pi L \simeq 3.3$. The finite volume effects  have been corrected by NLO ChPT\cite{GL2,CDH}. 
Fig.\ref{fig:nf2}(Right) shows $\mu a$ dependences of $a m_\pi^2/\mu$(left) and $af_\pi$(right) at $a=0.087$ fm, where $\mu = m_q\sin\theta_R $ is the twisted quark mass\cite{etmc2}.
Results indicate again that NLO ChPT fits work well at $m_\pi \le 500$ MeV for $m_\pi^2$ and $f_\pi$.
Fits including NNLO corrections or scaling violations have also been performed\cite{etmc3}.
It is then  found that the NNLO fits also works but the variation of the LEC $\ell_3$ is significant from NLO to NNLO while a change from $a=0.087$ to 0.067 fm seems to be explained by ChPT if the scaling violation is included at NLO.   

CERN-TorVergata groups have employed the non-perturbatively $O(a)$ improved Wilson quark action
for 2-flavor full QCD simulations at $a=0.052, 0.072, 0.078$ fm and $L=1.7\sim 1.9$ fm\cite{cern}. They have found that the quark mass dependence of  $m_\pi^2$ at $m_\pi = 377 \sim 495$ MeV is almost independent on the lattice spacing and is consistent with the NLO ChPT formula.

Let me summarize the current situation of the 2-flavor QCD and SU(2) ChPT.
First of all, the NLO ChPT describes lattice data of $m_\pi^2$ and $f_\pi$ well at $ m_\pi \le 500$ MeV and the expected continuum chiral-log is now unambiguously observed on the lattice for the first time in the unitary theory.
The NNLO ChPT may fit data beyond this pion mass, if some NLO LECs which starts appearing at NNLO are fixed to some phenomenological values. It turns out, however, that the NNLO corrections seems large, in particular for $f_\pi$, and that values of some NLO LECs are significantly affected.

I would like to give some remarks before closing this subsection.
In the 2-flavor QCD simulations mentioned above, possible finite size effects have been corrected by assuming that the ChPT formula for the corrections are valid.  Therefore it is important to check  this assumption by lattice simulations.  In table\ref{tab:FS},  magnitudes of finite size corrections, $R_O= \{O(L) - O(\infty\})/O(\infty)$ for $ O = m_\pi$ or $f_\pi$,  from lattice QCD simulation, the NLO ChPT\cite{GL2} and the resumed NLO ChPT\cite{CDH}, are compared in the case of the twisted mass QCD\cite{etmc4}. Although the resumed NLO ChPT formula is roughly consistent with lattice result, more detailed investigations are needed for the definite conclusion. 
To see whether the NNLO ChPT  indeed describes lattice data well, the NNLO ChPT fit {\it without} using phenomenological inputs should be tested. For this purpose, simultaneous fits to various quantities are needed to stabilize NNLO fits. For consistency, finite size corrections should be included in the fitting formula, instead of correcting lattice data before the fits.
Finally, from the theoretical point of view, inclusions of the lattice artifacts in ChPT should be made  to fits data correctly for Wilson-type quarks\cite{SS,aoki} including  twisted mass formulation\cite{AB}, as has been shown that it is mandatory to include such corrections in the case of the staggered-type quarks\cite{bernard}.

\begin{table}[bt]
\caption{Finite volume corrections to $m_\pi$ and $f_\pi$ from lattice QCD simulations\cite{etmc4}, NLO ChPT\cite{GL2} and resumed NLO ChPT formula\cite{CDH}}
\label{tab:FS}
\begin{center}
\begin{tabular}{||c||c|c|c||c|c|c||}
\hline
\hline
& \multicolumn{3}{| c ||}{ $R_{m_\pi} $\%} & \multicolumn{3}{| c ||}{ $R_{f_\pi} $\%} \\
\hline
$m_\pi L $ &  lattice & ChPT & resumed & lattice & ChPT & resumed \\
\hline
3.0 &  +6.2 & +1.8 & +4.7 & -10.7 & -7.3 & -8.9 \\
3.3 & +1.8 & + 0.62 & +1.0 & -2.5 & -2.5 & -2.4 \\
3.5 & +1.1 & +0.8 & +1.3 & -1.8 & -3.2 & -2.9 \\
\hline
\hline
\end{tabular}
\end{center}
\end{table}

\subsection{2+1 flavor QCD and ChPT}
The NLO formula of SU(3) ChPT for $m_\pi$ and $f_\pi$ are given by
\begin{eqnarray}
\frac{m_\pi^2}{m_l} &=& 2B_0 \left\{1+\mu_\pi -\frac{1}{3}\mu_\eta +\frac{2B_0}{f_0^2}\left[
16m_l (2L_8(\mu) - L_5(\mu)) + 16(2m_l + m_s)(2L_6(\mu)-L_4(\mu))\right]\right\} \nonumber\\
f_\pi &=& f_0 \left\{1-2\mu_\pi -\mu_K +\frac{2B_0}{f_0^2}\left[
8m_l  L_5(\mu) + 8(2m_l + m_s)L_4(\mu)\right]\right\} \\
\mu_{\rm PS} &=& \frac{\tilde m_{\rm PS}^2}{16\pi^2f_0^2} \ln \left(\frac{\tilde m_{\rm PS}^2}{\mu^2}\right), \quad \tilde m_\pi^2 = 2B_0 m_l, \ \tilde m_K^2 = B_0(m_l+m_s), \ \tilde m_\eta^2 = \frac{2B_0}{3}(m_l + m_s) \nonumber
\end{eqnarray}
where $B_0$ and $f_0$ are  LECs of SU(3) ChPT at LO, $L_i(\mu)$'s are those at NLO, $m_l$ is the degenerate up and down quark mass, and $m_s$ is the strange quark mass.
The validity of the NLO SU(3) ChPT for mass and decay constant of pion in the 2+1 flavor QCD, however,  is still unclear, since the strange quark mass is much heavier than up and down quark masses.  An alternative theoretical framework to describe  the quark mass dependence of these quantities is the SU(2) ChPT where the strange quark is treated as a heavy quark. The NLO formula for SU(2) ChPT are identical to eqs.(\ref{eq:mpi_su2}) and (\ref{eq:fpi_su2}) but
LECs  $B$, $f$, $\ell_{3,4}(\mu)$ appeared in the formula are $m_s$ dependent in this case. 

\begin{figure}[tb]
\centering
\includegraphics[width=65mm, angle=270,clip]{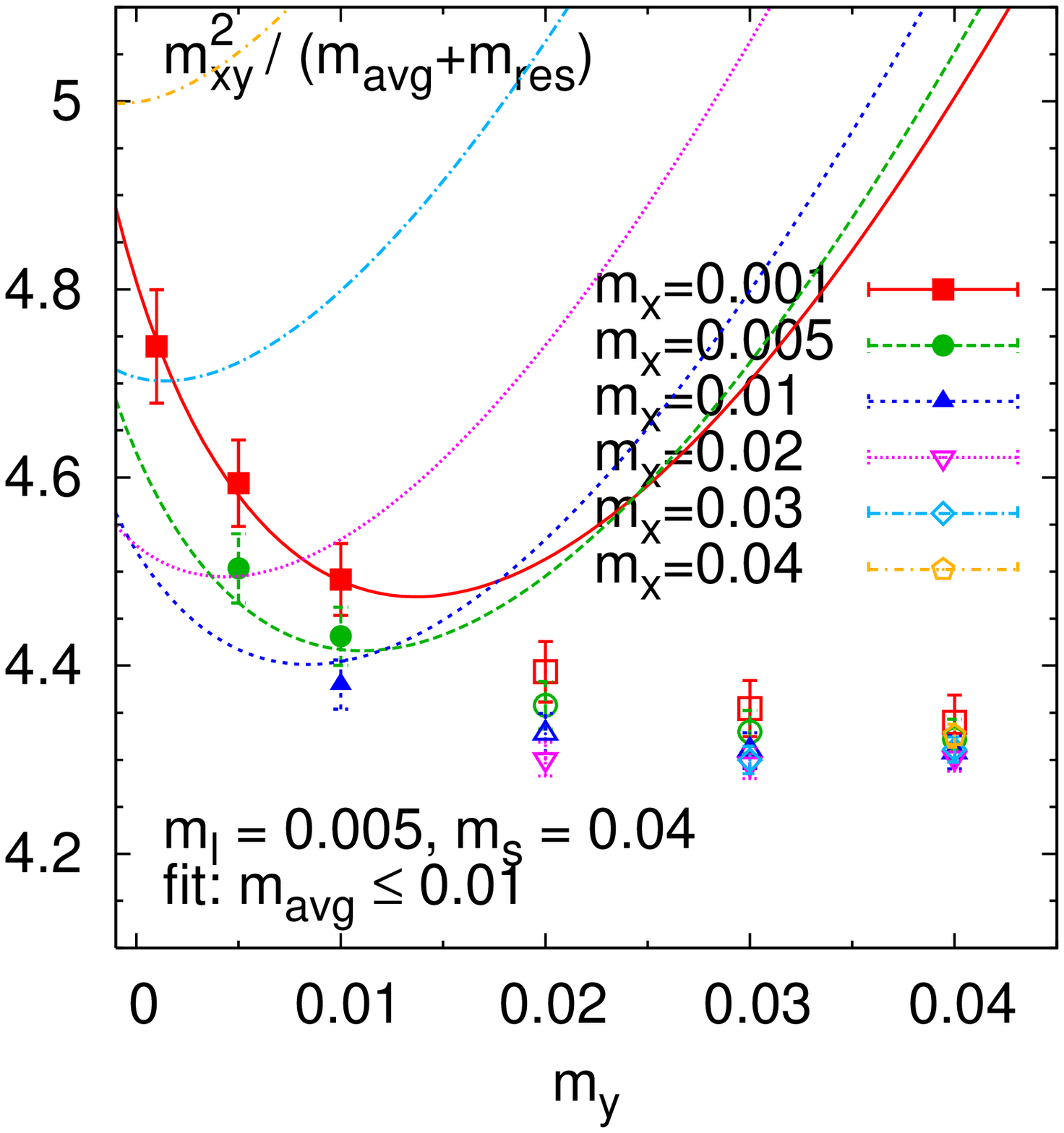}
\includegraphics[width=65mm, angle=270,clip]{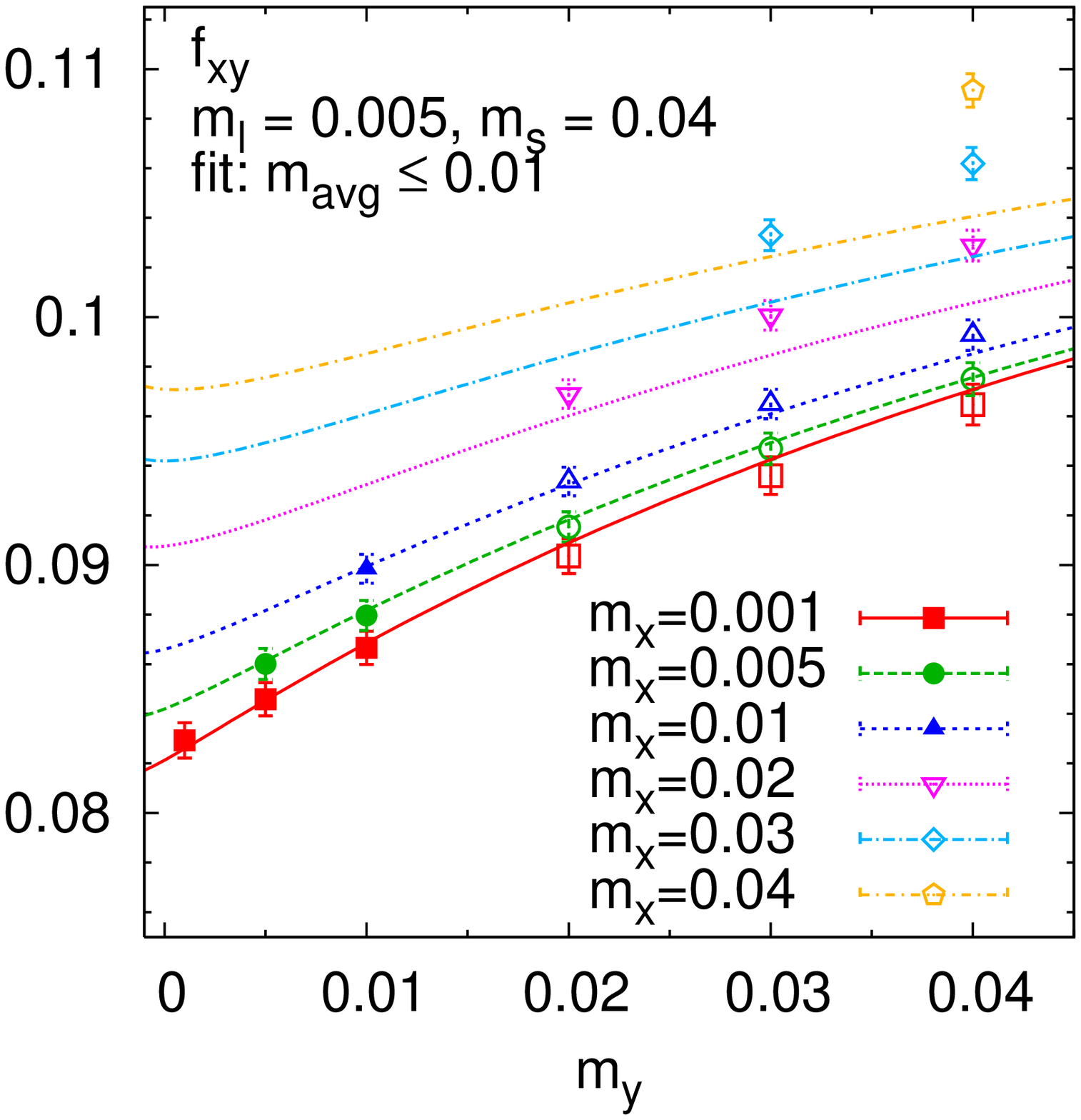}
\caption{Left: $m_{xy}^2/\tilde m_{\rm av}$ as a function of $m_y$ at $m_\pi^{\rm sea} = 331$ MeV in $N_f=2+1$ domain-wall QCD, together with NLO SU(2) PQChPT fit lines.
Only solid symbols are included in the fit. Right: Same  for $f_{xy}$. 
Both figures are taken from Ref.\cite{rbc-ukqcd}.}
\label{fig:rbc-ukqcd1}
\end{figure}

The SU(2) ChPT fit to the 2+1 flavor lattice QCD data has been first introduced by RBC-UKQCD Collaborations\cite{rbc-ukqcd}, and more recent analysis can be found in Ref.\cite{scholz}.
RBC-UKQCD Collaborations have performed simulations at $a=0.11$ fm with $L=2.7$ fm
and  at $a=0.08$ fm with $L=2.6$ fm,  where the minimum pion mass 330 MeV ($m_\pi L \simeq 4.6$) at $a=0.11$ fm and 310 MeV ($m_\pi L \simeq 4.1$) at $a=0.08$ fm.
They have employed the domain-wall quark whose additive mass renormalization term is small,
$m_{\rm res} a = 0.003$ at $a=0.11$ fm and 0.007 at $a=0.08$. Since the theory is almost chiral with this very small additive mass renormalization, the continuum ChPT at NLO is used for the analysis, with
the replacement that $m_f \rightarrow \tilde m_f = m_f + m_{\rm res}$.
They compare SU(2) and SU(3) partially quenched(PQ) ChPT at NLO, in order to describe $m_l$ dependences of mass and decay constant of pion, while the strange quark mass is fixed to one value in their simulations.
At $a=0.11$ fm, the partially quenched data of $m_\pi^2$ and $f_\pi$
are simultaneously fitted by the NLO SU(2) PQChPT at dynamical quark masses, $(m_la,m_sa)=(0.005,0.04)$ and $(0.01,0.04)$. 
Data and fitted lines are shown in Fig.\ref{fig:rbc-ukqcd1}, where $am_{xy}^2/(\tilde m_{\rm av})$ and $f_{xy}a$ are plotted as a function of $m_y a$ at $m_l a = 0.005$ ($m_\pi^{\rm sea} = 331$ MeV).  Here $m_{xy}$ is a mass of pseudo-scalar meson composed of two valence quarks whose masses are $m_x$ and $m_y$, and $m_{\rm av} = (m_x + m_y)/2$. The NLO SU(2) PQChPT fits data well at $ m_{\rm av} \le 0.01$(solid symbols) with $\chi^2/{\rm dof} \simeq 0.3$.
On the other hand, NLO SU(3) PQChPT works only at $m_{\rm av} \le 0.01$ with $\chi/{\rm dof} \simeq 0.7$, which can not cover the dynamical strange quark mass, $m_s = 0.04$, of this simulation.
Moreover the NLO correction at $m_\pi \simeq 500$ MeV in the unitary point( $m_l=m_x=m_y$) becomes 30-40\% for SU(2) and 60-70\% for SU(3),
which indicates that the NLO SU(3) ChPT is not sufficient for the strange quark.
The similar conclusion that the NLO SU(2) ChPT works better than the NLO SU(3) ChPT holds also at $a=0.08$ fm for $m_{\rm av} a \le 0.016$.

A similar conclusion that the SU(2) ChPT works much better than the SU(3) ChPT at NLO
is also obtained for non-chiral quark action by the PACS-CS collaboration\cite{pacs-cs,pacs-cs2,kuramashi}, who has employed non-perturbatively $O(a)$ improved Wilson quark action at $a=0.09$ fm. The absence of $O(a)$ scaling violations except small contributions  of $O( ma g^4)$ in the simulations justifies  the use of NLO ChPT formula of pion mass and decay constant even for this quark action\cite{ABIT,pacs-cs2}.
The SU(2) ChPT fits at NLO gives $\chi^2/{\rm dof} \simeq 0.4$ at $m_\pi < 500$ MeV. while the SU(3) fits leads to a 10-times larger value, $\chi^2/{\rm dof} \simeq 4$. 
In addition they have observed that the ratio of the NLO to the LO corrections  is much larger in SU(3) ChPT than in SU(2), suggesting worse convergence of the SU(3) ChPT at $m_\pi \simeq 500$ MeV.

The MILC collaboration employed the rooted staggered SU(3) PQChPT formula which include $O(a^2)$ lattice artifact, in order to fit pion mass and decay constant obtained with 2+1 flavor rooted staggered quarks at several lattice spacings. 
They have found\cite{milc,milc2} that the NLO SU(3) fit fails and  NNLO analytic terms (without log terms) are added to fit data at $m_x + m_y \le (0.39\sim 0.6) m_s$ where $m_x, m_y$ are valence quark masses and $m_s$ is the dynamical strange quark mass. This also suggests the failure of the NLO SU(3) PQChPT. Indeed, in this conference\cite{heller}, it is reported that
the NLO rooted staggered PQChPT with the NNLO continuum PQChPT work better for the SU(2) fit at fixed $m_s \simeq m_s^{\rm phys}$  and for the SU(3) fit at $m_s \le 0.6 m_s^{\rm phys}$ where
$m_s^{\rm phys}$ is the physical strange quark mass. If the NLO SU(3) ChPT works well  at $m_s$ much smaller than the physical value, one can determine the LECs of the NLO SU(3) ChPT more accurately than now.

Let me summarize the current status of the ChPT for the $N_f=2+1$ lattice QCD.
The NLO SU(3) (PQ)ChPT seems to fail at the physical strange quark mass while that of SU(2) seems to work at $m_\pi \le 500$ MeV. In the latter case, the strange quark mass dependence should be interpolated for physical predictions and LECs of the SU(2) ChPT.
In order to extract LECs of the SU(3) ChPT, it may be better to perform $N_f=3$ instead of $N_f = 2+ 1$   
simulations, keeping masses of all SU(3) Nambu-Goldstone boson smaller than 500 MeV.
A similar strategy is already taken for the staggered quark\cite{heller}, though the use of the chirally symmetric  quark action such as overlap action is preferable for this purpose. 
I would like to give one remark, made by Prof. J. Gasser during this conference, that
the NLO SU(3) ChPT may work at $m_K \le 500$ MeV, instead of $m_\pi \le 500$ MeV at fixed $m_s$.
To check this possibility, one must tune both $m_l$ and $m_s$ in order to keep the condition that
$m_K \le 500$ MeV. It will be interesting to perform such simulations.

\subsection{LECs and chiral behaviors} 
\begin{figure}[bt]
\centering
\includegraphics[width=55mm, angle=270,clip]{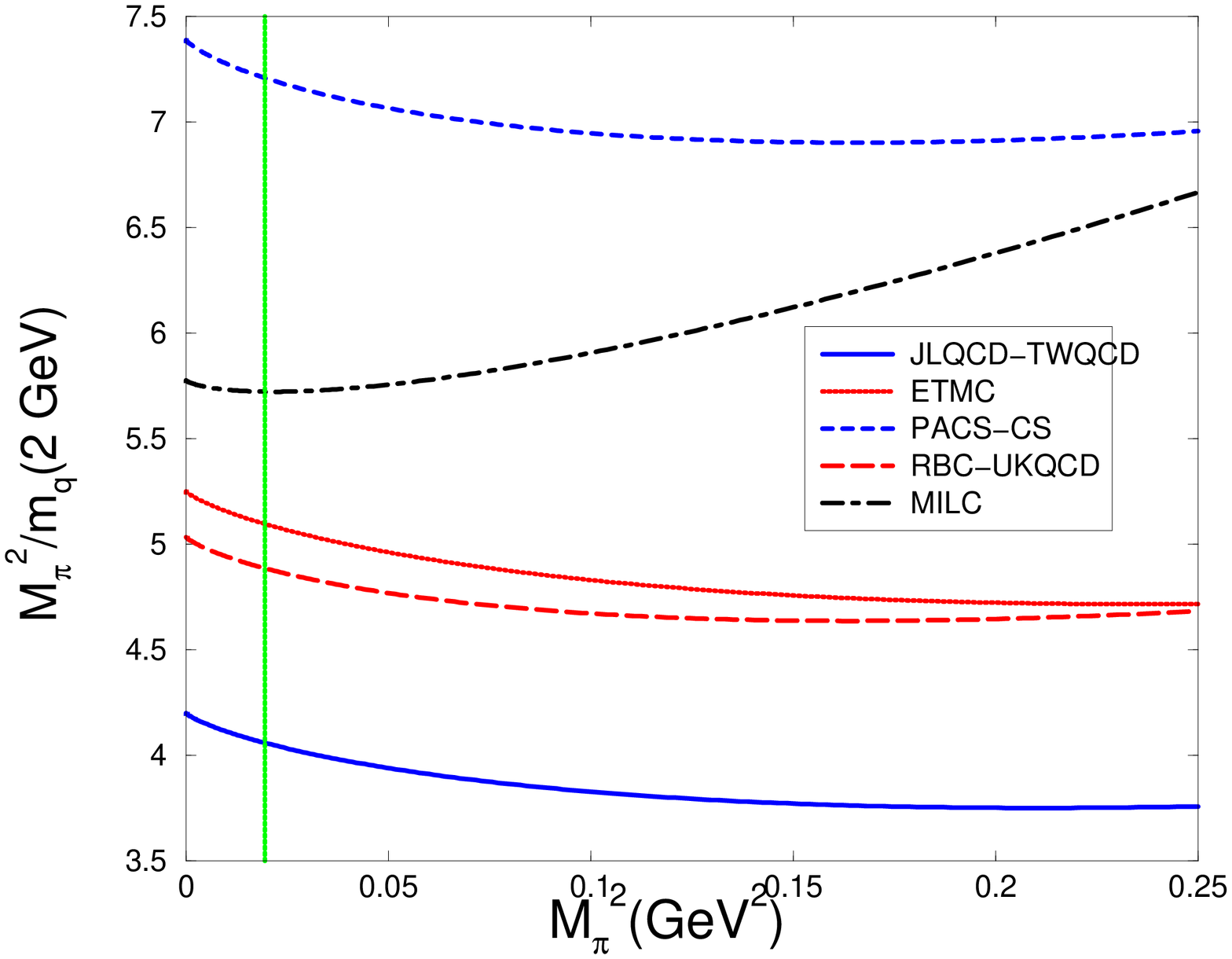}
\includegraphics[width=55mm, angle=270,clip]{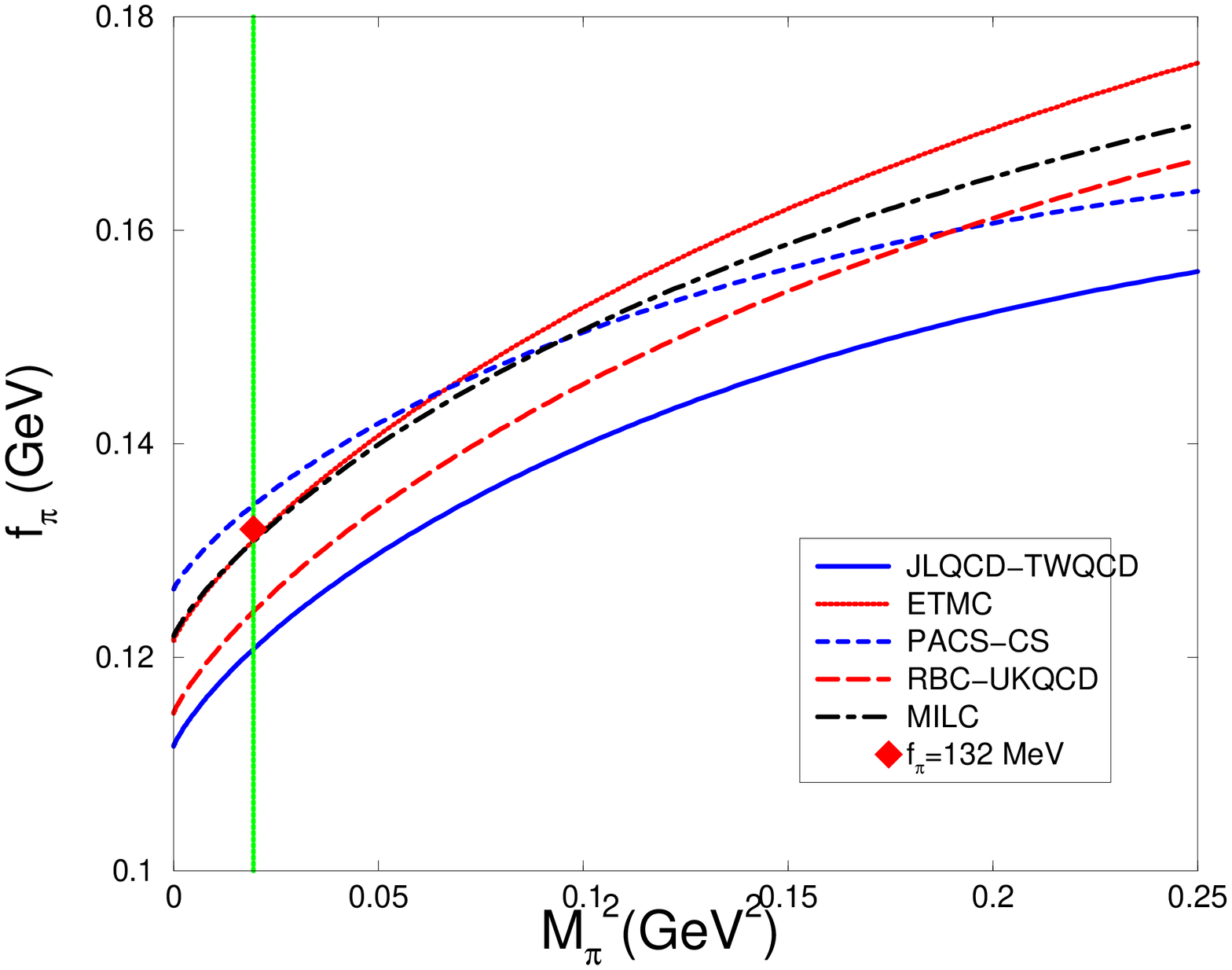}
\caption{$M_\pi^2/m_q$ (left) and $f_\pi$ (right) as a function of $M_\pi$ predicted by the NLO SU(2) ChPT formula, eqs. (\protect\ref{eq:mpi}) and (\protect\ref{eq:fpi}), for JLQCD-TWQCD(blue solid), ETMC(red dotted), PACS-CS(blue dashed), RBC-UKQCD(red long-dashed), MILC(black dot-dashed).
Green vertical lines indicate the physical point, $M_\pi = 139$ MeV. A red diamond gives $f_\pi=132$ MeV at the physical point. 
}
\label{fig:mass-dep}
\end{figure} 
Instead of comparing values of LECs $B$, $f$, $\ell_3$ and $\ell_4$ among various simulations,  in Fig.\ref{fig:mass-dep} we plot $M_\pi^2/m_q(2 {\rm GeV})$ and $f_\pi$ as a function of $M_\pi^2$,
which is given by
\begin{eqnarray}
\frac{M_\pi^2}{m_q(2 {\rm GeV})} &=& 2B ({\rm 2GeV}) \left\{ 1+ \frac{M_\pi^2}{16\pi^2f^2}
\left[\ln \left(\frac{M_\pi^2}{m_\pi^2}\right) -\ell_3(m_\pi)\right] \right\}, 
\label{eq:mpi}  \\
f_\pi &=& f \left\{ 1- \frac{M_\pi^2}{8\pi^2f^2}
\left[\ln \left(\frac{M_\pi^2}{m_\pi^2}\right) -\ell_4(m_\pi)\right] \right\},
\label{eq:fpi}
\end{eqnarray}
where we take $\mu = m_\pi = 139$ MeV.
Data are taken from both $N_f=2$(JLQCD-TWQCD\cite{jlqcd_nf2} and ETMC\cite{etmc5}) and $N_f=2+1$(MILC\cite{milc}, RBC-UKQD\cite{rbc} and PACS-CS\cite{pacs-cs}) simulations.
Although results among different groups more or less converge for $f_\pi$, sizable differences are observed for $M_\pi^2/m_q$. One of reasons for these differences seems to come from the overall renormalization factor of $m_q$:
MILC and PACS-CS collaborations, who employ perturbative renormalization factors, tend to give larger values of the ratio than those from
JLQCD-TWQCD, ETMC and RBC-UKQCD collaborations, who use non-perturbative estimates for the renormalization.   

\section{Pion form factors}
In this section, we consider form factors of the pion, 
where chiral perturbation theory plays an important role to make extrapolations in terms of the momentum transfer as well as the quark mass.
The vector and scalar form factors are defined by
\begin{eqnarray}
\langle \pi (p^\prime)\vert V_\mu \vert \pi(p)\rangle &=&
(p+p^\prime)_\mu F_V(q^2), \quad 
\langle \pi (p^\prime)\vert S \vert \pi(p)\rangle = F_S(q^2),
\quad q^2=(p-p^\prime)^2 \\
\langle r^2\rangle_X &=& 6\left. \frac{\partial F_X(q^2)}{\partial q^2}\right\vert_{q^2=0},
\quad
c_X = \left. \frac{\partial^2 F_X(q^2)}{\partial (q^2)^2}\right\vert_{q^2=0},
\quad X=V,S
\end{eqnarray}
where $V_\mu$ ($S$) are the vector current (scalar density) in QCD, and
$\langle r^2\rangle_V$ ( $\langle r^2\rangle_S$ ) is called the charge(scalar) radius.
Recent full QCD calculations for the pion form factors are summarized in table \ref{tab:PFF}.

\begin{table}[tb]
\caption{(Recent full QCD calculations for pion form factors. Here $Q^2 = -q^2$.}
\label{tab:PFF}
\begin{center}
\begin{tabular}{|c|c|c|c|c|c|c|c|}
\hline
\hline
Group & $N_f$ & quarks & $a$(fm) & $L$(fm) & $Q^2({\rm GeV}^2)$  &
 $m_\pi$(MeV) & $ F_X$ \\
\hline
\hline
QCDSF-UKQCD\cite{qcdsf-ukqcdFF} & 2& clover & 0.07-0.12 & 1.4-2.0
& 0.31 -4.3 & 400-1000 & V \\
RBC-UKQCD\cite{rbc-ukqcdFF} & 2+1 & DW & 0.11 & 2.8 & 0.013-0.258 & 330 & V \\
ETMC\cite{etmcFF} & 2 & TM & 0.07-0.09 & 2.2-2.9 & 0.05-0.8 & 260-580 & V \\
JLQCD-TWQCD\cite{jlqcdFF} & 2 & overlap & 0.12 & 1.9 & 0.252-1.7 & 290-750 & V,S \\
\hline
\hline
\end{tabular}
\end{center}
\end{table}

\begin{figure}[tb]
\centering
\includegraphics[width=50mm, angle=270,clip]{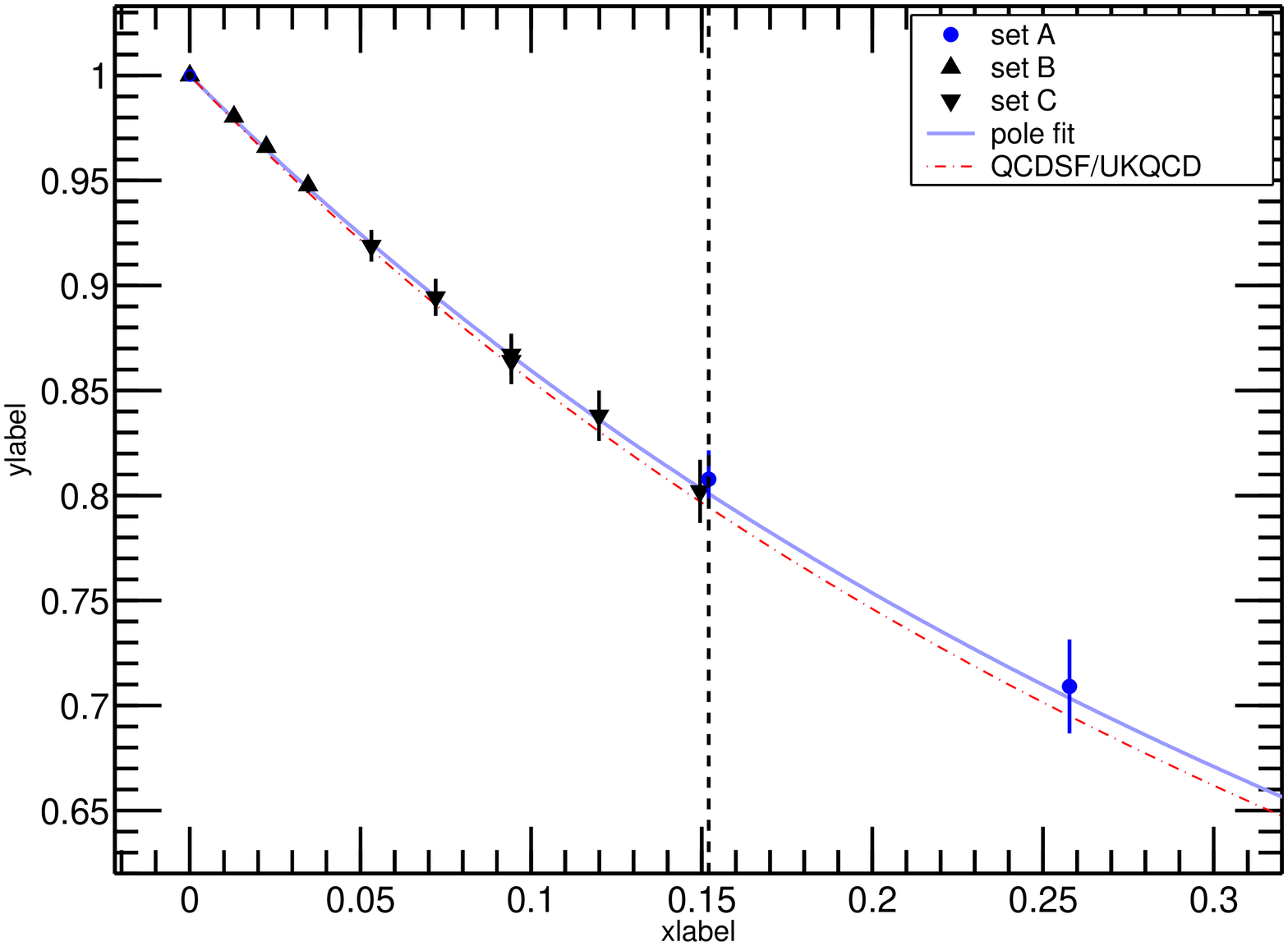}
\includegraphics[width=50mm, angle=270,clip]{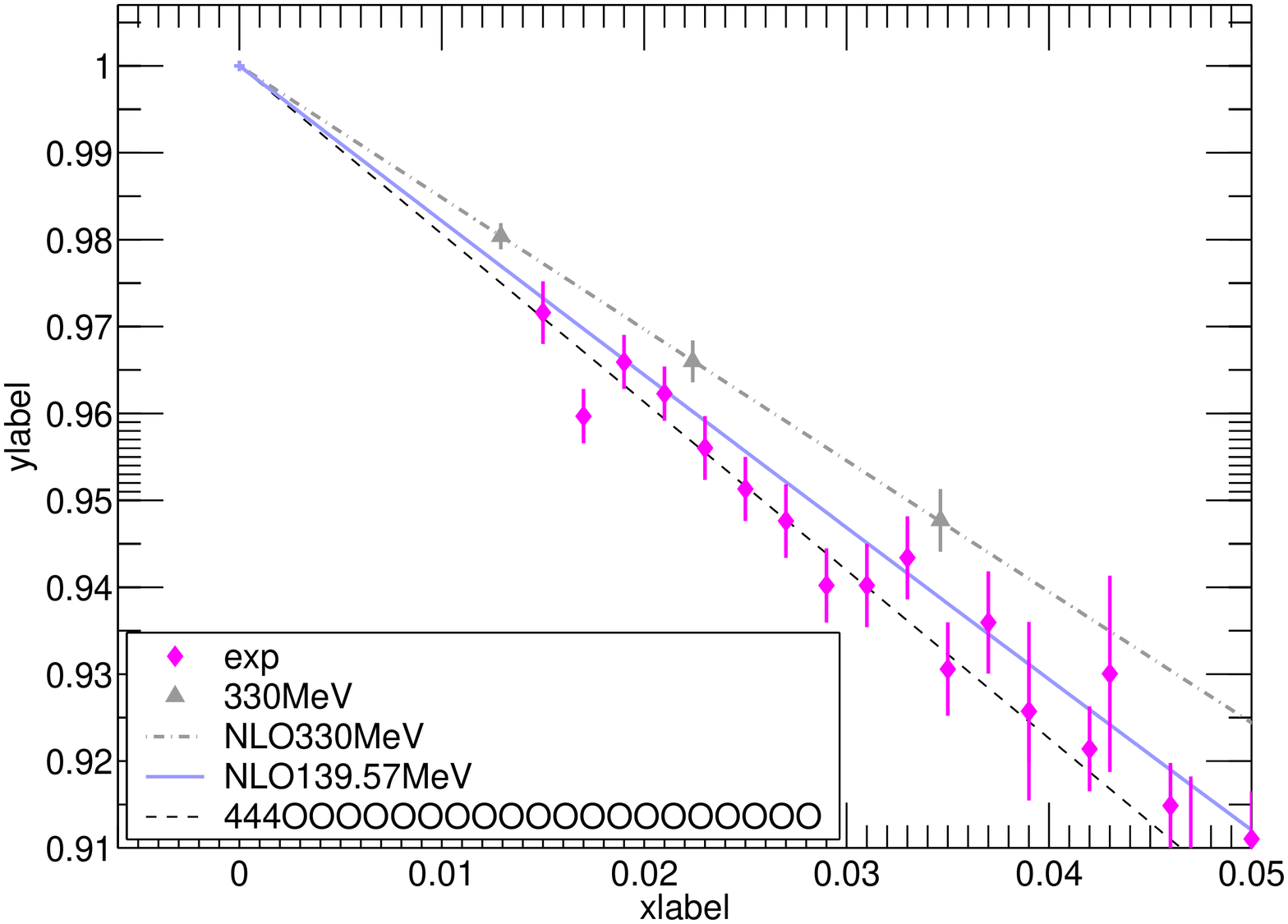}
\caption{(Left) The vector form factor of pion $F_V(q^2)$ as a function of $Q^2 = - q^2$, together with the single pole fit (solid line) in $N_f=2+1$ flavor domain-wall QCD at $a=0.11$ fm and $m_\pi^2 = 330$ MeV\cite{rbc-ukqcdFF}. The dotted line is the single pole fit of Ref.\cite{qcdsf-ukqcdFF} at the same pion mass but in $N_f=2$ QCD with clover quarks.
(Right)  The NLO SU(2) ChPT fit at  $m_\pi^2 = 330$ MeV (dash-dotted line), $139.57$ MeV (Solid line) and with the charged radius being fixed to the PDG world average(dashed line) for  small $Q^2$, together with lattice data at  $m_\pi^2 = 330$ MeV(triangles)  and experimental data (diamonds). 
Both figures are taken from Ref.\cite{rbc-ukqcdFF}.}
\label{fig:rbc-ukqcdFF}
\end{figure}

It has been  found by all groups that $q^2$ dependence of $F_V(q^2)$ at fixed quark mass is well described by the pole ansatz:
\begin{eqnarray}
F_V(q^2) &=& \frac{1}{1-q^2/M_{\rm Pole}^2},
\end{eqnarray}
in particular at small $q^2$, as seen in Fig.\ref{fig:rbc-ukqcdFF}(Left).
Moreover the value of $M_{\rm pole}$ obtained by the fit is closed to the vector meson mass $M_\rho$ at this quark mass (Vector Meson Dominance).
The single pole ansatz leads to
\begin{eqnarray}
\langle r^2\rangle_V \simeq \frac{6}{M_{\rm pole}^2}, \quad c_V \simeq \frac{1}{M_{\rm pole}^4}
\simeq \left(\frac{\langle r^2\rangle_V}{6}\right)^2  &\Rightarrow&
\langle r^2\rangle_V \simeq 6\sqrt{c_V} .
\end{eqnarray}
On the other hand, the SU(2) ChPT at NLO gives
\begin{eqnarray}
\langle r^2\rangle_V^{\rm NLO} &=& -\frac{2}{(4\pi f)^2}\left\{ 1+ 6 N \ell_6^r(\mu) + \ln[m_\pi^2/\mu^2] \right\}, \quad
c_V^{\rm NLO} = \frac{1}{30(4\pi f m_\pi)^2},
\end{eqnarray}
which, together with the relation from the single pole ansatz, predicts
\begin{eqnarray}
\langle r^2\rangle_V \simeq \sqrt{\frac{6}{5}}\times \frac{1}{4\pi f m_\pi} \simeq 0.22 {\rm fm}^2,
\end{eqnarray}
which is far below the experimental value, $\langle r^2\rangle_V^{\rm exp. PDG} = 0.452(11)$ fm$^2$. 
This discrepancy mainly comes from the fact that the NLO ChPT does not reproduce behavior of lattice data and a single pole ansatz.

Because of this problem, QCDSF-UKQCD collaborations\cite{qcdsf-ukqcdFF} give up the ChPT fit and employ the single pole ansatz for the $q^2$ dependence of $F_V(q^2)$. After the chiral extrapolation to the physical point  by the form that $M_{\rm pole}^2 = c_0 + c_1 m_\pi^2$, they obtain   
$\langle r^2\rangle_V = 0.441(19)(56)(-29)$ fm$^2$, which agrees well with the experimental value.
The form factor $F_V(q^2)$ itself, the single pole ansatz extrapolated to the physical pion mass, reproduces experimental data well.

Regardless of the success of the single pole fit,  the situation is not satisfactory from the theoretical point of view, since the NLO ChPT is incompatible with it.

RBC-UKQCD collaborations\cite{rbc-ukqcdFF} have fitted $F_V(q^2)$ at very small $q^2$, instead of $\langle r^2\rangle_V$ and $c_V$, by the NLO SU(2) ChPT formula given as
\begin{eqnarray}
F_V(q^2) &=& 1 + \frac{1}{f^2}\left[ - 2\ell_6^r(\mu)q^2 + 4 {\cal H}(m_\pi^2, q^2,\mu^2)\right] \\
{\cal H}(x,y,z) &=& \frac{ x H(y/x)}{32\pi^2}-\frac{y}{192\pi^2}\ln \frac{x}{z}, \quad
H(x) = -\frac{4}{3}+\frac{5x}{18}-\frac{x-4}{6}\sqrt{\frac{x-4}{x}}\ln\left(\frac{\sqrt{(x-4)/x}+1}{\sqrt{(x-4)/x}-1}\right) . \nonumber
\end{eqnarray}
Using $f a = 0.0665(47)$ from the fit of the decay constant as an input, one unknown LEC $\ell_6^r$  can be extracted form one point  at $q^2 = -0.013$ GeV$^2$ as
$\ell_6^2(m_\rho) = -0.093(10)$, though no degree of freedom is left.
This value of the LEC leads to $\langle r^2 \rangle_V = 0.354(31)$ fm$^2$ at the simulation point where $m_\pi = 330$ MeV, which becomes $\langle r^2 \rangle_V = 0.418(38)$ fm$^2$ at the physical point, $m_\pi = 139$ MeV. The form factors $F_V(q^2)$ reconstructed by the NLO SU(2) ChPT formula with the $\ell_6^2$ are compared with lattice data at $m_\pi = 330$ MeV and with experimental values at $m_\pi=130$ MeV in Fig. \ref{fig:rbc-ukqcdFF}, which shows reasonable agreements in both cases.

The ETM collaboration\cite{etmcFF} has performed a little more advanced analysis, employing the NNLO SU(2) ChPT formula for the form factor. 
Using the experimental value of $\langle r^2 \rangle_S = 0.61(4)$ fm$^2$ to fix some LECs through the NNLO formula, they have simultaneously fitted $m_\pi$, $f_\pi$ and $F_V(q^2)$ by the NNLO SU(2) ChPT formula at $ 0.05 \le Q^2 \le 0.8$ ( GeV$^2$), which predicts
$\langle r^2 \rangle_V = 0.438(29)$ fm$^2$ at $m_\pi = 139$ MeV.
An agreement in $F_V(q^2)$ between ChPT fit results at NNLO and experimental data is reasonably
good.

\begin{figure}[tb]
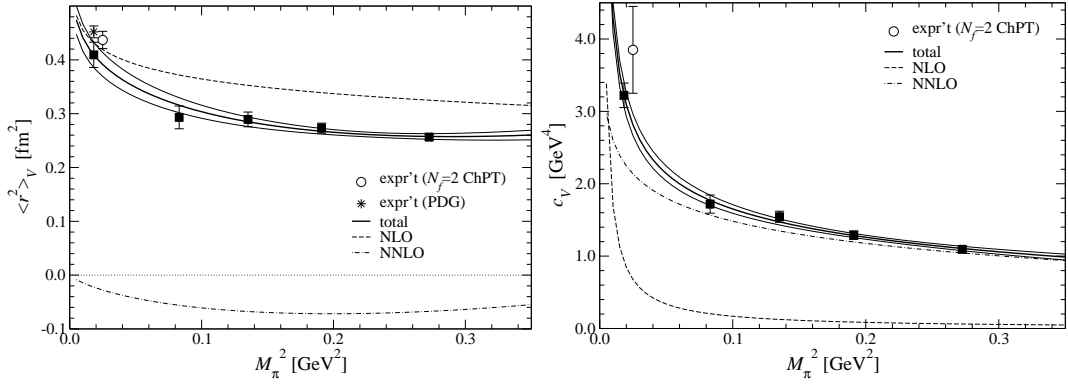

\centering
\includegraphics[width=70mm, angle=0,clip]{Figs/JLQCD_PFF_r2V.eps}
\includegraphics[width=70mm, angle=0,clip]{Figs/JLQCD_PFF_cV.eps}
\caption{Left: The charge radius $\langle r^2 \rangle_V $(solid squires) as a function of $M_\pi^2$, together with the NNLO SU(2) ChPT fit(solid line),  its NLO (dot-dashed line) and  NNLO (dashed line) contributions. The open circle is the experimental value extracted by ChPT and the asterisk  is the PDG value. Right: The same for $c_V$.}
\label{fig:jlqcdFF}
\end{figure}

JLQCD-TWQCD collaborations\cite{jlqcdFF} have used a hybrid method for the fit:
$q^2$ dependences of $F_V$ and $F_S$ are fitted by the single pole plus polynomials in $q^2$, from which $\langle r^2 \rangle_V$, $c_V$ and $\langle r^2 \rangle_S$ are extracted.
On the other hand, as shown in Fig.\ref{fig:jlqcdFF}, $m_\pi^2$ dependences of these three quantities are fitted by the NNLO SU(2) ChPT formula, which give
$\langle r^2 \rangle_V = 0.409(23)$ fm$^2$, $c_V =3.22(17)$ GeV$^{-4}$($\simeq 0.0049$ fm$^4$), and $\langle r^2 \rangle_S = 0.617(79)$ fm$^2$. It is interesting to note that $c_V$ turns out to be very close to a value predicted from this value of $\langle r^2 \rangle_V $ by the single pole ansatz,
$c_V \simeq 0.0047$ fm$^4$, though the convergence of the expansion is questionable from the fact that $c_V^{\rm NLO} < c_V^{\rm NNLO}$ ( Note that $c_V^{\rm LO}= 0$. ).
It remains  an important challenge  to fit both $q^2$ and $m_\pi^2$ dependence of $F_X(q^2)$ by the NNLO SU(2) ChPT formula. 

\section{Summary}
Let me conclude this review. Now chiral-logs of the NLO ChPT for mass and decay constant of pion
are clearly seen in lattice QCD even for data in unitary theories.
Recent analysis suggest that the NLO SU(2) ChPT works for these quantities at pion mass less than 500 MeV, while an applicability of the NLO SU(3) ChPT to the physical strange quark mass seems questionable, so that the NLO SU(2) ChPT formula is often used even for data in 2+1 flavor QCD simulations.
As far as pion form factors are concerned, more detailed investigations are needed for a comparison between numerical data and ChPT, in particular, for the convergence of the chiral expansion.
In future more complicated quantities from lattice QCD, such as the $\pi\pi$ scattering phase shift, should be compared with the ChPT predictions.

\end{document}